\documentclass[final,5p,times,twocolumn]{elsarticle}
\usepackage{subfigure}
\usepackage{amssymb}
\usepackage{lipsum}
\usepackage{amsthm}
\usepackage{graphicx}
\usepackage{amsmath}
\usepackage[colorlinks=true, linkcolor=blue, citecolor=blue, urlcolor=blue]{hyperref}
\usepackage{mathrsfs}

\journal{ }

\begin{document}

\begin{frontmatter}



\title{Kiselev black hole and the ultra-slow evaporating behavior}


\author[first]{Chen-Hao Wu}
\ead{chenhao\_wu@nuaa.edu.cn}
\author[first,second]{Xiao Liang}
\ead{xliang@nuaa.edu.cn}
\author[first,third]{Ya-Peng Hu}
\ead{huyp@nuaa.edu.cn(Corresponding author)}
\affiliation[first]{organization={College of Physics, Nanjing University of Aeronautics and Astronautics}, 
city={Nanjing},
postcode={210016},
country={China}}
\affiliation[second]{organization={Nanjing No.13 High School},
city={Nanjing},
postcode={210008},
country={China}}
\affiliation[third]{organization={MIIT Key Laboratory of Aerospace Information Materials and Physics,  Nanjing University of Aeronautics and Astronautics},
city={Nanjing},
postcode={210016},
country={China}}

\begin{abstract}
Kiselev solution is a metric that describes black holes immersed in a quintessence-like dark energy background. By introducing a dynamic state parameter $w_q$, the Kiselev solution is supposed to help comprehend the effect of quintessential matter on black holes. In this work, we study the evaporation behaviors of Kiselev black holes. By varying the state parameter $w_q$, we find that the decreasing state parameter lowers the non-final stage temperature and markedly prolongs the evaporation lifetime. We also find that the ultra-slow evaporation mechanism of Kiselev black holes differs vastly from the perfect fluid dark matter (PFDM) black holes and Horndeski black holes, which share the analogous ultra-long lifetime. These results illuminate the effects of dynamic dark energy background on black hole evaporation, provide a potential laboratory to constrain the value of $w_q$, and may complement cosmological and astrophysical observations, e.g., the DESI’s preference for thawing dark energy and the observation of exploding black holes based on ultra-slow evaporation.
\end{abstract}



\begin{keyword}
black hole evaporation \sep Kiselev solution \sep quintessence-like background


\end{keyword}

\end{frontmatter}




\section{Introduction}
Over a quarter of a century ago, astrophysical observations revealed the accelerating expansion of our universe \cite{SupernovaCosmologyProject:1998vns, SupernovaSearchTeam:1998fmf}. This phenomenon has also been confirmed by a growing number of independent observations \cite{WMAP:2003elm, Planck:2015fie, Planck:2018vyg, Brout:2022vxf, DES:2024jxu, DESI:2024mwx}. To explain this phenomenon, some physicists introduced the concept of \textit{dark energy}, which means that our universe contains some elements with negative pressure \cite{Copeland:2006wr, Peebles:2002gy}. However, the origin of dark energy remains a puzzle, and no single theory has been able to satisfy all self-consistency requirements. In the past 20 years, the $\Lambda$CDM model has been in an unshakable position. In such a model, a constant $\Lambda$ describes dark energy, and most observations are compatible with this model \cite{Peebles:2002gy, Weinberg:1988cp, Padmanabhan:2002ji}. Recently, the $\Lambda$CDM model has faced increasing challenges from various astrophysical and cosmological tests \cite{Perivolaropoulos:2021jda}. Notably, the Dark Energy Spectroscopic Instrument (DESI) first-year result recently shows a preference for thawing dark energy \cite{DESI:2024mwx}. As a well-known thawing dark energy model, \textit{quintessence} theory was first raised to introduce evolving dark energy with a time-evolving equation of state \cite{Caldwell:1997ii}, which is trusted to be one possible interpretation of the accelerating expansion of our universe at late times. 

Considering the quintessential dark energy filling of our universe, it is natural to extend this background to compact objects, such as black holes. The most well-known attempt was made by Kiselev, also known as the Kiselev solution \cite{Kiselev:2002dx}. This solution describes spherically symmetric black holes surrounded by anisotropic fluids. It is believed that it can effectively provide a dark energy background for black hole solutions if the appropriate parameters are selected. Here we must point out that the Kiselev solution is not an exact black hole solution with ``quintessence'': It cannot provide a precise isotropic pressure to satisfy the perfect fluid spacetime required by the quintessence model \footnote{Embedding black holes in perfect fluids is generally difficult, but recent research indicates that such anisotropic background might provide a possible explanation for the mass growth of some supermassive black holes \cite{Wu:2026xwo}.}, just as the comments from Ref.\cite{Visser:2019brz}. Despite some controversies, we can still argue that it provides an effective solution as a quintessential matter mimicker. The qualities of Kiselev black holes (KBHs) have been extensively discussed in Refs.\cite{Toshmatov:2015npp, Chen:2008ra, Zeng:2020vsj, Li:2014ixn, Fernando:2012ue, Chen:2005qh, Xu:2016jod, Wu:2018meo, Wang:2019kzk, Pugliese:2024rlr, Chen:2024qlm, Lungu:2024iob}, however, researchers have yet to address some certain crucial thermodynamic properties, specifically the evaporation (evolution) of black holes in such a frame.

The evaporation of black holes is an intriguing topic in black hole physics. It was derived from black hole radiation, also called Hawking radiation now. By applying quantum field theory in curved spacetime, Hawking proved that black holes can emit thermal particles \cite{Hawking:1974rv, Hawking:1975vcx}. Then two years later, Page was the first to apply this theory to Schwarzschild black holes and found that the radiation spectrum is consistent with the graybody spectrum \cite{Page:1976df}. One can find that a Schwarzschild black hole with initial mass $M_0$ will keep losing its mass (which means nothing is left!) until its lifetime undergoes $t \sim M_0^3$. This result raises another mystery of physics, namely the information paradox (This is beyond the scope of this paper, and one can see more discussion in Refs.\cite{Almheiri:2020cfm, Chen:2014jwq}). Besides the uncharged, nonrotating case, Page also extended his investigation to the Kerr black hole \cite{Page:1976ki}. In 1990, by combining the Schwinger effect in quantum electrodynamics, Hiscock and Weems tried their efforts to extend such a study to the charged case \cite{Hiscock:1990ex}. In this case, the lifetime of black holes is largely prolonged. Except for these ``classic'' cases, the research on the evaporation behavior of black holes has already become a field teeming with abundant achievements: diverse black hole cases were tested in various backgrounds. To name a few, Ref.\cite{Page:2015rxa} discusses the black hole evaporation in anti-de Sitter spacetime, whereas Ref.\cite{Gregory:2021ozs} investigates the de Sitter case. Numerous modified gravity theories, dark energy, and dark matter models provide fertile ground for testing the evaporative behavior of black holes \cite{Ong:2018syk, Xu:2018liy, Xu:2019krv, Xu:2020xsl, Hou:2020yni, Wu:2021zyl, Liang:2023jrj, An:2024fzf, Liang:2024ygf}. In light of the above progress, it is necessary to extend the study of black hole evaporation behavior immersed in a quintessence-like background.

This paper is organized as follows: In Sect.(\ref{2}), we first review the Kiselev solution and its thermodynamics. We also present its asymptotic behavior compared with the Schwarzschild case. Then, we investigate the evaporation behavior of the KBH and give numerical results in Sect.(\ref{3}). Finally, in Sect.(\ref{4}), we conclude the main text with a summary and discuss the meaning of KBH evaporation behavior for the future investigation of time-evolving dark energy.
Throughout this paper, we use the natural units with $\hbar=k_B=G =c=1$.

\section{Thermodynamics of Kiselev black holes}\label{2}
In this section, we will first introduce the KBH solutions. Generally, the role of electric charge in astrophysics is usually negligible in consideration of the neutralization of the galactic medium. Therefore, we consider a static and spherically symmetric black hole solution and assume that the ansatz takes the following form
\begin{eqnarray}
  ds^2=-f(r)dt^2+\frac{1}{f(r)}dr^2+r^2(d\theta^2+\sin^2\theta d\varphi^2),\label{action}
\end{eqnarray}
which satisfies the Einstein field equation
\begin{eqnarray}
R_{\mu\nu}-\frac{1}{2}R g_{\mu\nu}=8\pi T^{\rm DE}_{\mu\nu},
\end{eqnarray}
where $T^{\rm DE}_{\mu\nu}$ is the stress-energy tensor of the quintessential matter, denoted as
\begin{eqnarray}
\begin{aligned}
&{T^{\rm DE}}^{t}_{~t}={T^{\rm DE}}^r_{~r}=-\rho_q, \\
&{T^{\rm DE}}^{\theta}_{~\theta}={T^{\rm DE}}^{\varphi}_{~\varphi}=\frac{1}{2}\rho_q\left(3w_q+1\right).\label{emt}
\end{aligned}
\end{eqnarray}
Here $\rho_q$ is the density of the quintessential matter expressed as
\begin{eqnarray}
\rho_q=-\frac{a}{2}\frac{3w_q}{r^{3(w_q+1)}},
\end{eqnarray}
in which $a$ is the normalization factor connecting with the cosmological observation \footnote{Note that the factor $a$ is proportional to the quintessence density, which implies the value of $a$ should be very small, referring to the actual observation. Here, we adopt the statement given in the literature \cite{Zeng:2020vsj}, and the normalization factor $a=0.05$ is taken as we aim to simplify the calculations and emphasize the asymptotic behavior of black hole thermodynamics. The current actual observation value is lower than $10^{-21}$ times this amount.}, and one can easily find that the parameter $w_q$ describes the equation of state when defining the average pressure
\begin{eqnarray}
\bar{p}:=\frac{{T^{\rm DE}}^r_{~r}+{T^{\rm DE}}^{\theta}_{~\theta}+{T^{\rm DE}}^{\varphi}_{~\varphi}}{3}=w_q \,\rho_q.
\end{eqnarray}

When the state parameter $w_q$ is restricted to the range of $-1 < w_q < -1/3$, this model is believed to help mimic the quintessential matter surrounding the black holes. From the above equations, one can easily find the KBH solution, derived as \cite{Kiselev:2002dx}
\begin{eqnarray}
  f(r)=1-\frac{2M}{r}-\frac{a}{r^{3w_q+1}},
  \label{metric}
\end{eqnarray}
where the parameter $M$ represents the mass of the black hole. When $a=0$, this metric reduces to the Schwarzschild black hole. In the case of
$w_q \rightarrow -1$, the black hole solution will evolve to the Schwarzschild-de Sitter (SdS) case, and that implies this solution has two event horizons. The presence of the exponential term precludes an analytical solution for the horizon radius $r_+$ by solving the relation $f(r_+)=0$. However, we can find a special case to obtain a set of analytical horizon radii by setting $w_q=-2/3$. The two roots are written as
\begin{eqnarray}
  r_+^{\rm BH}=\frac{1-\sqrt{1-8 a M}}{2 a}, \, r_+^{\rm C}=\frac{1+\sqrt{1-8 a M}}{2 a},
\end{eqnarray}
in which case, the restricted condition $8 a M <1$ is preset. Here, the small one represents the black hole's event horizon, and the large one is the cosmological horizon. It is worth noting that the cosmological horizon can also emit the particles \cite{Gibbons:1977mu}; however, the thermodynamics of SdS spacetime is still a thorny question, let alone such a more complicated case (one can refer to the Refs.\cite{Gregory:2021ozs, Qiu:2019qgp} to see more details). Consequently, we only consider the radiation from the black hole and ignore the cosmological horizon part, given that the cosmological parameters from the experiment \cite{Planck:2018vyg} imply the radiation from the cosmological horizon is very weak. 

When the event horizon of the black hole is determined by $f(r_+)=0$ (since we ignore the radiation from the cosmological horizon, we will then use $r_+$ to represent the event horizon radius of the black hole hereafter), we can now express the mass of a black hole in terms of the radius of its event horizon as
\begin{eqnarray}
 M=\frac{1}{2}\left(r_+ -ar_+^{-3w_q}\right) ,\label{mass}
\end{eqnarray}
and then the Hawking temperature and  entropy of the black hole can be obtained as
\begin{eqnarray}
\begin{aligned}
&T=\left.\frac{f'(r)}{4\pi}\right|_{r=r_+}=\frac{1}{4 \pi}\left(\frac{1}{r_+}+\frac{3aw_q}{r_{+}^{2+3w_q}}\right),\label{eq:6}\\
&S=\int^{r_+}_{0}\frac{1}{T}\left(\frac{\partial M}{\partial r_+}\right)d r_+=\pi r_+^2.\label{eq:7}
\end{aligned}
\end{eqnarray}

\begin{figure}[h]
    \centering
    \includegraphics[width=0.95\linewidth]{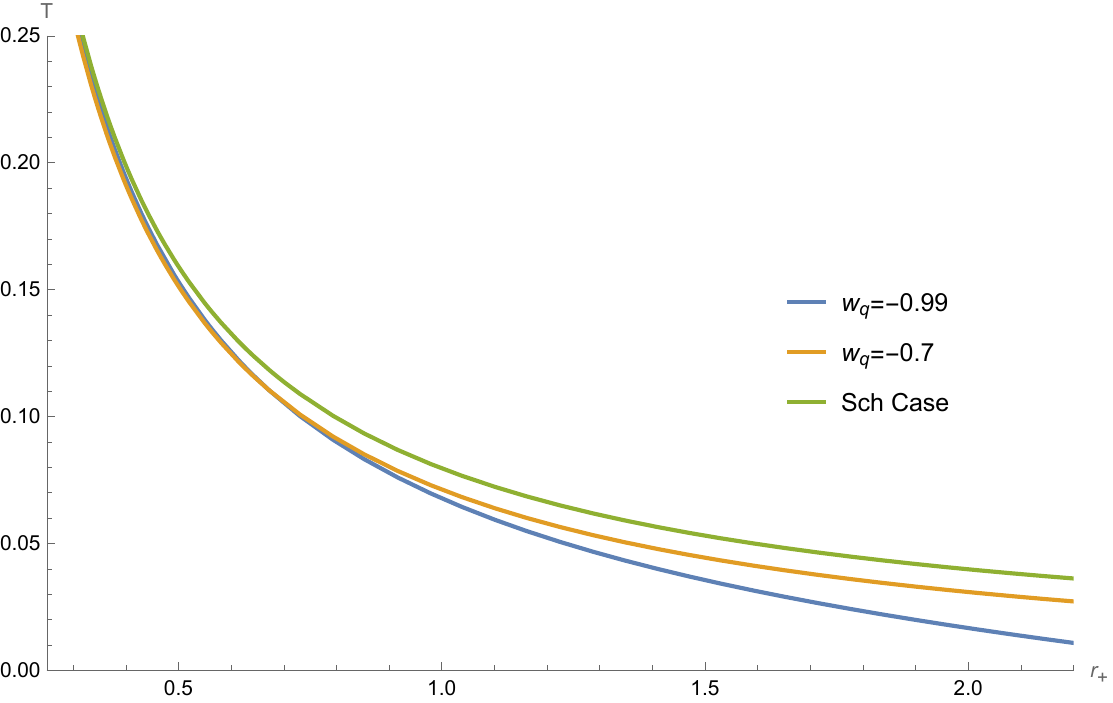}
    \caption{The temperature of KBHs with two benchmarks of state parameter (with $a=0.05$) versus the Schwarzschild case.} 
    \label{Tplot}
\end{figure}

We plot the temperature of KBHs with two benchmarks of the state parameter, compared with the Schwarzschild case, in FIG.\ref{Tplot}. One can find that the temperature of the black hole is monotonically decreasing in the possible mass ranges, which implies the evaporation behavior of the KBH may be consistent with the Schwarzschild cases. Although in terms of general temperature behavior, the KBHs and Schwarzschild black holes share an analogous profile, the details are not the same. On the right half of FIG.\ref{Tplot}, the KBHs with a smaller state parameter tend to a lower temperature, which implies that, except for the last stage of evaporation, the state parameter will effectively influence the evaporation rate. We will discuss in detail how the state parameters affect the asymptotic behavior of black hole evaporation in the next section.

\section{The evaporation behavior of the Kiselev black hole}\label{3}
Having established the thermodynamic properties of the KBH, we now turn to its dynamical evolution. In this section, we first compute the asymptotic evaporation rate and then provide specific lifespan estimations.
\subsection{The asymptotic behavior}
Since the black hole behaves like a real thermal system, we can now explore its evolution behavior by calculating the emission rate of energy.

Complying with Page's calculation, we know the thermal radiation part from the black holes follows the Stefan-Boltzmann law \cite{Page:1976df, Xu:2019wak}
\begin{eqnarray}
 \frac{dM}{dt} = -a_d \sigma T^4,\label{sbl}
\end{eqnarray}
where $a_d= \pi^2/15$ is the radiation density constant, and the parameter $\sigma$ is the cross section of the emission source connecting with the species of the emitted particles. Here we take the geometrical optics approximation (the high-frequency limit) from Ref.\cite{Page:1976df} because the particles with long wavelengths are difficult to radiate to the outside. Considering that the thermal radiation from black holes mostly consists of massless particles, e.g., photons, neutrinos, and gravitons, the cross section can now be expressed as $\sigma=\pi b_c^2$, where $b_c$ is the critical impact factor we will discuss later.

To find out the specific forms of $b_c$, we shall investigate the null geodesics in the KBH spacetime. We start from the standard Euler–Lagrange equation
\begin{eqnarray}
	\frac{d}{d\lambda}\frac{\partial \mathscr{L}}{\partial \dot{x}^\mu}-\frac{\partial \mathscr{L}}{\partial x^\mu}=0,
\end{eqnarray}
where $\dot{x}^\mu$ is the derivative of coordinates to the affine parameter, namely $\dot{x}^\mu=dx^\mu/d\lambda$. Considering the spherical symmetry of spacetime, we can take particles moving in the equatorial plane without loss of generality, i.e., $\theta=\pi/2$, thus the Lagrangian of the system can be expressed as
\begin{eqnarray}
\mathscr{L}=\frac{1}{2}g_{\mu\nu}\dot{x}^\mu\dot{x}^\nu=\frac{1}{2}\left(  -f(r)\dot{t}^2+f(r)^{-1}\dot{r}^2+r^2\dot{\phi}^2\right) .
\end{eqnarray}
Again, given the symmetry of spacetime, the metric $g_{\mu\nu}$ is independent of $t$, $\phi$, we have
\begin{eqnarray}
\frac{d}{d\lambda}\frac{\partial \mathscr{L}}{\partial \dot{t}}=0,\,\,
\frac{d}{d\lambda}\frac{\partial \mathscr{L}}{\partial \dot{\phi}}=0.
\end{eqnarray}

We can now define two conserved quantities for time and azimuthal direction as
\begin{eqnarray}
\begin{aligned}
E&=-g_{tt}\dot{t}=f(r)\dot{t},\\
L&=g_{\phi\phi}\dot{\phi}=r^2\dot{\phi},\label{EL}
\end{aligned}
\end{eqnarray}
which represents the energy conservation and angular momentum conservation of particles, respectively. Now defining $\kappa:=-g_{\mu\nu}\dot{x}^\mu\dot{x}^\nu$, and the $\kappa=0$ case corresponds to the null geodesics. Substituting the two conserved quantities of Eq.\eqref{EL} into $\kappa=0$ and shifting the affine parameter to $\tilde{\lambda}=L\lambda$, we can find the geodesic equation
\begin{eqnarray}
\left({\frac{d r}{d \tilde{\lambda}}}\right)^2=\frac{1}{b^2}-V_{\rm eff}(r),
\end{eqnarray}
where $b:=L/E$ is the impact factor, and $V_{\rm eff}(r):=f(r)/r^2$ represents the effective potential of particles. One can find that massless particles can reach infinity only if
\begin{eqnarray}
\frac{1}{b^2}\geq V_{\rm eff}(r),
\end{eqnarray}
for all $r>r_+$. Thus, we can now define the critical impact factor as
\begin{eqnarray}
b_c:=\frac{r_p}{\sqrt{f(r_p)}}
\end{eqnarray}
by solving the extremum condition $\partial_r V_{\rm eff}(r)|_{r=r_p}=0$, where $r_p$ is the radius of the unstable circular orbit of massless particles consistent with the photon sphere. 

Considering the effective potential $V_{\rm eff}(r)=f(r)/r^2$ depends on the exact form of metric function $f(r)$, it still contains the complicated exponential term when $w_q \neq -2/3$. That means we cannot obtain the analytic roots of the photon sphere radius and its corresponding critical impact factor, except for the special case $w_q=-2/3$ with 
\begin{eqnarray}
b_c=\frac{\sqrt{3}(1-\sqrt{1-6 a M })}{a \sqrt{-1+2\sqrt{1-6 a M }}}.
\end{eqnarray}

\begin{figure}[h]
    \centering
    \includegraphics[width=0.95\linewidth]{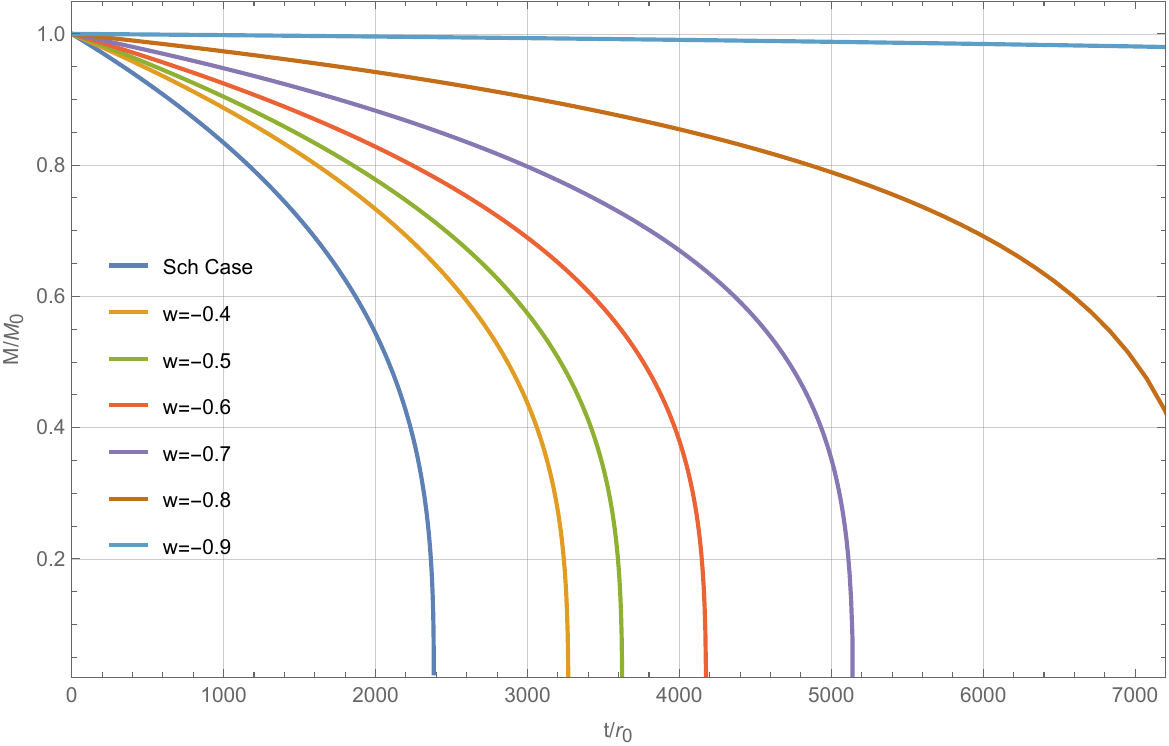}
    \caption{The evolution of mass $M$ during the evaporation process in KBHs with different initial state parameters (with $a=0.05$). We also present the Schwarzschild black hole for comparison.} 
    \label{Mplot}
\end{figure}

In light of the above, we apply the numerical method to obtain asymptotic evaporation behavior with different state parameters. We select different state parameters and plot these cases on the same graph, see FIG.\ref{Mplot}. We can observe that the state parameters $w_q$ have a significant impact on the black hole evaporation process. When $w_q \rightarrow -1/3$, the evaporation behavior of the KBH shares a similar feature with the Schwarzschild case. The black hole tends to lose its mass relatively rapidly. That is reasonable because the term of metric $a/r^{3w_q+1} \rightarrow a$ when $w_q \rightarrow -1/3$, and the metric Eq.\eqref{metric} tends to reduce to the Schwarzschild type considering the value of $a \ll 1$. When the state parameters gradually decrease, the lifetime of the KBHs gradually increases. 

The evaporation behavior shown in FIG.\ref{Mplot} is consistent with the qualitative expectation from the Stefan–Boltzmann law \eqref{sbl}, where the emission rate is highly sensitive to the temperature through the $T^4$ dependence. In the limit of the geometrical optics approximation, the photon energy term $T^4$ is of higher order, making the asymptotic behavior of the temperature $T$ the dominant factor in the mass decay rate. As shown in FIG.\ref{Tplot}, for smaller $w_q$, the black hole temperature is closer to the extremal state compared to cases with larger $w_q$ or the Schwarzschild case. Consequently, a smaller $w_q$ corresponds to a more pronounced suppression of thermal emission during most of the evaporation process. This prolonged lifespan is reminiscent of the analogous ultra-slowly evaporating behavior observed in perfect fluid dark matter (PFDM) black holes \cite{Liang:2023jrj} and Horndeski black holes \cite{Liang:2024ygf}. However, the underlying physical mechanisms vary significantly. For the PFDM black holes, the lifetime is extended via two channels: the reduction of the effective surface gravity, and the energy loss through the generation of dark matter particles via a dark Schwinger effect (i.e., $dM/dt \sim - (c_1 T^4 + c_2 dQ_d/dt)$), similar to the Reissner-Nordström (RN) black hole \cite{Hiscock:1990ex}. In the Horndeski model, the ultra-slow evaporation is driven by non-minimally coupled matter, modifying the emission channels as $dM/dt \sim - (c_1 T^4 + c_2 dQ/dt + c_3 dQ_\phi/dt)$ and leading to a three-stage evaporation rate, which is very different from the RN case. In contrast to these models, the slow evaporation in the Kiselev case originates purely from the modification of the thermal emission temperature induced by the quintessential background, without invoking additional particle production channels.

To quantitatively understand why the evaporation lifetime is significantly prolonged for small $w_q$, we shall analyze the exact scaling behavior of the emission rate. By factoring out $1/r_+$ in the temperature expression (see Eq.\eqref{eq:6}), the Hawking temperature can be rewritten as
\begin{eqnarray}
T(r_+) = \frac{1}{4\pi r_+} \left[ 1 - 3 a |w_q| r_+^{-(1+3w_q)} \right] .
\end{eqnarray}
For quintessential dark energy ($-1 < w_q < -1/3$), the exponent $-(1+3w_q)$ is strictly positive. As $w_q \to -1$, this exponent approaches $2$. For a macroscopic black hole with a large initial horizon radius $r_+$, the term $r_+^{-(1+3w_q)}$ acts as a powerful amplifier for the quintessential correction, causing a strong cancellation against the leading term $1$ and drastically lowering the temperature. According to the Stefan–Boltzmann law, the mass loss rate is proportional to the emission cross section $\sigma \sim r_+^2$ and $T^4$:
\begin{eqnarray}
\frac{dM}{dt} \propto - \sigma T^4 \sim - \frac{1}{r_+^2} \left[ 1 - 3 a |w_q| r_+^{-(1+3w_q)} \right]^4 .
\end{eqnarray}
Due to the highly non-linear fourth-power dependence, even a modest temperature drop triggered by the quintessential cancellation results in a severe suppression of the evaporation rate. When integrating to obtain the lifetime $\tau \sim \int (dt/dM) dM$, the suppressed rate $(1 - 3 a |w_q| r_+^{-(1+3w_q)})^{-4}$ forces the time integral to increase rapidly as $w_q$ decreases. Therefore, the ultra-slow evaporation is not a mere phenomenological coincidence, but mathematically originates from the $T^4$ amplification of the power-law cancellation encoded in the metric.

\subsection{Concrete examples}
In the preceding subsection, we have described how the quintessence-like matter affects the asymptotic behavior of the KBH evaporation. In this subsection, we present selected examples and plug in the specific cosmological parameters to quantify this effect.

We adopt the estimation method from Page's literature \cite{Page:1976df}, and we only take the massless particles (three kinds of neutrinos with their antiparticles, photons, and gravitons with two polarizations each) into account to simplify the calculation. The total power of the emitted particles would be
\begin{equation}
\begin{aligned}
 \frac{dM}{dt} = 
 -a_d T^4
[\frac{7}{16}\sigma(\nu_{e})
+ \frac{7}{16}\sigma(\bar{\nu}_{e})
+ \frac{7}{16}\sigma(\nu_{\mu})
+ \frac{7}{16}\sigma(\bar{\nu}_{\mu})\\
+ \frac{7}{16}\sigma(\nu_{\tau})
+ \frac{7}{16}\sigma(\bar{\nu}_{\tau})
+ \sigma(\gamma)
+ \sigma(g)]\label{eqm}
\end{aligned}
\end{equation}
for emission of neutrinos $\nu_{e}, \nu_{\mu}, \nu_{\tau}$, photons $\gamma$ and gravitons $g$. By applying the geometrical optics approximation, the cross sections of all particles go to $27\pi M^2$ determined by the photon sphere $r_p=3M$ for a Schwarzschild black hole, and the emission power becomes $\frac{333}{163840 \pi} M^{-2}$. Just as Page pointed out, such a power estimation is slightly larger than the numerical result because we obviously overestimated the emission power of photons and gravitons. However, given that we omit other standard model particles, the estimation of the black hole lifetime would be fixed to a proper scale. For a Schwarzschild black hole with mass $ M_i=5.4 \times 10^{14}\, \text{g} \simeq 2.481 \times 10^{19}$ (a value assumed in many studies, below which Primordial Black Holes (PBHs) would have exploded by today) in the natural units, the evaporation lifetime will be estimated as
\begin{eqnarray}
\tau_{\rm Sch} \approx 4.24266 \times 10^{17}\, \text{s} \approx 13.4534\, \text{Gyr},
\end{eqnarray}
which is very close to the most widely accepted age of the universe $\tau_{\rm U} \approx 4.35\times 10^{17} \, \text{s} \approx 13.8 \,\text{Gyr}$. In a word, the above processing is an effective estimation for black hole evaporation, and we will take this mass as our benchmark in the following examples.

Now we consider a black hole immersed in time-evolved quintessence-like matter, and such a picture should be distinguished from examples involving the cosmological constant. One may notice that in the asymptotic behavior part, we have set the normalization factor $a$ to $0.05$. Obviously, this value will be much smaller than this amount in actual observations. In the subsequent calculations, we will set this value to $\mathcal{O}(10^{-40})$, which value is not too big but makes the density of quintessence-like matter $\rho_{q}$ much greater than the density of $\rho_{\Lambda} \sim 10^{-122}$. The first concrete example is the $w_q=-2/3$ in view of its well-analytical form. For convenience, we plug the relation $M=\frac{1}{2}r_+(1-ar_+)$ into Eq.\eqref{eqm}, and the power can now write as
\begin{eqnarray}
\frac{d r_+}{d t}=\frac{37\,(-2 a + 1/r_+)^4\,\bigl(-1 + \sqrt{1 - 3 a r_+ + 3 a^2 r_+^2}\bigr)^2}%
{5120\,a^2\,\pi\,(1 - 2 a r_+)\,\bigl(-1 + 2 \sqrt{1 - 3 a r_+ + 3 a^2 r_+^2}\bigr)}.
\end{eqnarray}
For a black hole with the same initial mass $M_i$, we get a lifespan as
\begin{eqnarray}
\tau_{ w_q \sim -\frac{2}{3}} \approx 4.24267 \times 10^{17}\, \text{s} \approx 13.4534\, \text{Gyr},
\end{eqnarray}
almost identical to the Schwarzschild case. This is foreseeable because the black hole temperature $T= (r_+^{-1}-2a)/4\pi \sim 1/(4\pi r_+)$ when $1/r_+ \gg a$, and the event horizon initial radius $r_i^{-1} \sim \mathcal{O}(10^{-19})$ is obviously larger than the value of $a$. 

When the universe evolves, it is natural to consider the situation of cosmological accelerating expansion, which is consistent with the strengthening of dark energy. To formulate a general quantitative conclusion, we numerically calculated the evaporation lifetime for a continuous spectrum of state parameters $w_q\in(-1,-1/3)$ with a fixed small normalization factor $a=10^{-40}$. The numerical results are summarized in Table \ref{table}. Interestingly, the data reveal a sharp transition in the evaporation behavior; for $w_q>-0.9$, the quintessential energy correction becomes subdominant, and the lifetime remains virtually indistinguishable from the Schwarzschild baseline $\tau_{\rm Sch} \approx 13.4534\, \text{Gyr}$. However, as $w_q$ approaches the de Sitter limit ($w_q \rightarrow -1$), the negative thermal correction is exponentially amplified, leading to a drastic extension of the black hole's lifespan.

\begin{table}[htbp!]
\centering
\begin{tabular}{|c|c|}
\hline
$ w_q $ &  Lifetime (Gyr) \\
\hline
-0.99&	20.7658\\
-0.98&	14.7867\\
-0.97&	13.7727\\
-0.96&	13.534\\
-0.95&	13.474\\
-0.9&	13.4534\\
-0.8&	13.4534\\
-0.7&	13.4534\\
-0.6&	13.4534\\
-0.5&	13.4534\\
-0.4&	13.4534\\
\hline
\end{tabular}
\caption{The evaporation lifetime of KBH with initial mass $ M_i=5.4 \times 10^{14}\, \text{g}$ versus different state parameters $w_q$, and we set $a=10^{-40}$.}
\label{table}
\end{table}

Based on these numerical results, we deduced an \textit{empirical formula} to characterize the relationship between $w_q$ and the evaporation lifetime in the quintessential-dominated regime
\begin{eqnarray}
\tau(w_q) \approx 13.4534 + 39.2428 \,e^{-168.057(1+w_q)} \quad \text{Gyr}
\end{eqnarray}
This formula quantitatively captures the transition from standard Hawking evaporation to the ultra-slow evaporation regime. The large decay constant ($\sim -168$) in the exponential term implies that the ultra-slow mechanism is highly sensitive to the proximity of the dark energy equation of state to the cosmological constant limit.

This result largely extends the evaporation lifetime compared with the Schwarzschild case. Due to our limited understanding of the detailed evolution of the universe, it is difficult to give a precise picture to describe such a process. However, one can easily imagine an image in which a black hole was produced during the early stage of the universe, and it underwent a long process of Schwarzschild-like evaporation. But in late times, the strength of dark energy increased rapidly, making the behavior of evaporation diverge from the Schwarzschild-like case, and leaving us an intriguing open question: whether the lower limit of the PBHs mass constrained by Hawking evaporation increases due to this mechanism?

\section{Conclusion and discussion}\label{4} 
In this work, we firstly reviewed the thermodynamics of KBHs based on the field equation and the stress-energy tensor of quintessential matter. As shown in FIG.\ref{Tplot}, the temperature of KBHs exhibits a monotonically decreasing trend with increasing black hole mass, consistent with the general thermal behavior of Schwarzschild black holes. However, significant differences emerge in detail: for large event horizon radii, KBHs with smaller state parameters $w_q$ tend to have lower temperatures. This indicates that the quintessential background modulates the thermal emission intensity of black holes, laying the foundation for understanding their evaporation behavior. Then, we explored the evaporation dynamics of KBHs by applying Page’s geometrical optics approximation. Since the state parameter appears in the exponent of $r$, making it difficult to obtain an analytical form for the critical impact factor, we adopted numerical methods to simulate the mass evolution. Our results reveal that the state parameter $w_q$ is a dominant factor controlling the evaporation lifetime: As $w_q$ decreases, the metric term $a/r^{3w_q+1}$ evolves into $a r^2$ and becomes the dominant component of the metric. This leads to a significant reduction in the thermal emission rate, resulting in an ultra-slow evaporation process. We clarified that this ultra-slow evaporation mechanism originates from the thermal emission correction induced by the quintessential background. This differs fundamentally from the other ultra-slow evaporation black hole models we have investigated, such as PFDM black holes \cite{Liang:2023jrj} (which rely on reduced effective surface gravity and dark Schwinger pair production) and Horndeski black holes \cite{Liang:2024ygf} (driven by non-minimally coupled effects), highlighting the unique effects of quintessence-like backgrounds.

Lastly, we discussed the possibility that this mechanism could broaden the upper limit of the mass of PBHs. These findings have twofold implications for cosmology and astrophysics. The evaporation characteristics of KBHs could be used to constrain the value of $w_q$, complementing existing cosmological observations: one can easily recall the DESI’s preference for thawing dark energy. For such dark energy research, KBHs provide a ``laboratory" to probe a dynamic state parameter. Recently, there has been a novel viewpoint suggesting that the ultra-slow evaporation will make it highly likely to observe an exploding black hole over the next 10 years \cite{Baker:2025zxm}. Different from the dark-QED toy model mentioned in the literature, we here introduce a different mechanism that can also result in ultra-slow evaporation. Considering ubiquitous dark energy distributed throughout the universe, we have reasons to believe that such a mechanism would lead to similar outcomes.

\section*{Acknowledgements}
This work is supported by the National Natural Science Foundation of China (NSFC) under Grant No.12175105. 


\bibliographystyle{elsarticle-num} 
\biboptions{sort&compress}
\bibliography{example}

\end{document}